%% file: verifinisq-ExtAbs.tex
\documentclass[a4paper,10pt,twoside,reqno,nonamelimits]{article}
\usepackage{xr}
\usepackage{a4wide}
\usepackage[colorlinks=true,urlcolor=blue,linkcolor=blue,citecolor=blue]{hyperref}
\usepackage{newcent}         % PSNFSS

\usepackage{amsmath}
\usepackage{amssymb}
\usepackage{amsfonts}
\usepackage{amscd}
\usepackage{amsthm}
\usepackage{amsxtra}
\usepackage{mathrsfs}
\usepackage{nicefrac}
\usepackage{upgreek}
\usepackage{graphicx}%\graphicspath{ {image/} }
\usepackage{xcolor}
\usepackage{subcaption}
\usepackage{tikz} \usetikzlibrary{arrows.meta, positioning, fit, calc}

\usepackage{enumitem}
\setlist[enumerate]{leftmargin=2em,itemindent=0em, labelindent=0pt,labelwidth=1.5em,labelsep=.5em, align=left, noitemsep}
\newlist{condenum}{enumerate}{1}
\setlist[condenum]{ leftmargin=4em,itemindent=0em, labelindent=0pt,labelwidth=1.5em,labelsep=.5em, align=left, noitemsep}
\newlist{txtenum}{enumerate}{1}
\setlist[txtenum]{leftmargin=0em,itemindent=1.5em, labelindent=0pt,labelwidth=1em,labelsep=.5em, align=left}
\usepackage{algorithm}
\usepackage{algpseudocodex}
\algrenewcommand{\algorithmicfor}{\textbf{For}}
\algrenewcommand{\algorithmicif}{\textbf{If}}

\usepackage{datetime}

\input{defsN.amsart}
\input{defs.common}
\usepackage{titlesec}
\renewcommand{\thesubsubsection}{\thesubsection.\alph{subsubsection}}
\renewcommand{\theparagraph}{\thesubsubsection.\roman{paragraph}}
\renewcommand{\thesubparagraph}{\theparagraph\arabic{subparagraph}:}
\titleformat{\paragraph}[runin]{\normalfont\bfseries\color{pink!10!black!70}}{\theparagraph}{1em}{}[]
\titleformat{\subparagraph}[runin]{\normalfont\slshape}{\thesubparagraph}{1em}{}[]
\titlespacing*{\paragraph}{0pt}{1ex}{1em}
\titlespacing*{\subparagraph}{0pt}{1ex}{1em}

\newcommand{\myparsmall}{\par\smallskip\noindent}
\newcommand{\mypar}{\par\medskip\noindent}

\usepackage{diagbox}

\newcommand{\mycaption}[2]{\caption{\textbf{#1 }\small{#2}}}
\newcommand{\good}{\texttt{\tiny good}}
\newcommand{\bad}{\texttt{\tiny bad}}
\newcommand{\trip}[2]{\tr(#1\,#2)}
\begin{document}
\title{Certifying a stabilizer state with few observables but many shots}%
\author{Dirk Oliver Theis
  \\[1ex]
  \small Theoretical Computer Science, University of Tartu, Estonia\\
  \small \texttt{dotheis@ut.ee}%
}
%%%%%%%%%%%%%%%%%%%%%%%%%%%%%%%%%%%%%%%%%%%%%%%%%%%%%%%%%%%%%%%%%%%%%%%%%%%%%%%%%%%%%%%%%%%%%%%%%%%%%%%%%%%%%%%%%%%%%%%%%%%%%%%%%%%%%%%%%%%%%%%%%%%%%%
%%
%%
\date{Fri Jul 18 14:18:45 CEST 2025
  \\
  Compiled: \currenttime}
%%
%%
%%%%%%%%%%%%%%%%%%%%%%%%%%%%%%%%%%%%%%%%%%%%%%%%%%%%%%%%%%%%%%%%%%%%%%%%%%%%%%%%%%%%%%%%%%%%%%%%%%%%%%%%%%%%%%%%%%%%%%%%%%%%%%%%%%%%%%%%%%%%%%%%%%%%%%
\maketitle

%%%%%%%%%%%%%%%%%%%%%%%%%%%%%%%%%%%%%%%%%%%%%%%%%%%%%%%%%%%%%%%%%%%%%%%%%%%%%%%%%%%%%%%%%%%%%%%%%%%%%%%%%
%%% arXiv abstract:
%
%
%
%
%
%%%%%%%%%%%%%%%%%%%%%%%%%%%%%%%%%%%%%%%%%%%%%%%%%%%%%%%%%%%%%%%%%%%%%%%%%%%%%%%%%%%%%%%%%%%%%%%%%%%%%%%%%
\begin{abstract}
  We propose a quantum-state-certification protocol for stabilizer states, motivated by application in in-situ
  testing of NISQ-era quantum computer systems: The number of qubits is bounded, and in terms of cost of running
  the protocol, identical repetition of quantum circuits contribute negligibly compared to switching the
  measurement bases.

  The method builds on Direct Fidelity Estimation and work by Somma et al.~(2006), but replaces linear averages by
  a minimum over estimates of expectation values.
  We provide mathematically rigorous analysis of the false-negative and false-positive rates.

  \par\medskip%

  \par\medskip\noindent%
  This is an extended abstract, the full paper is available at\\\texttt{dojt.srht.site/storage/2412.16690}.
\end{abstract}

\setcounter{secnumdepth}{4}

\begin{center}\footnotesize
\setcounter{tocdepth}{2}
\tableofcontents
\end{center}

\section{Introduction}\label{sec:intro}
%% Brief description
%%%% Topic: QSC
This paper proposes a novel approach to \emph{Quantum State Certification} (QSC) for stabilizer states.

%%%% DEF: q.s. *Certification*
We use the term QSC to refer to the task of checking, through measurements, that a state that can be
experimentally prepared is either close to or far away from a known target state that's supposed to be
prepared.

%%%% *Math* Certification
Concretely, we assume two parameters $0\le \delta<\eps \le 1$, and the requirement that the certification
method, with overwhelming probability, accepts any ``good'' state~$\varrho$ with fidelity $\ge 1-\delta$
with~$\varrho_0$ and rejects any ``bad'' state with fidelity $\le 1-\eps$.
(In this paper, for brevity, we will write ``fidelity'' instead of ``fidelity against the target state'', as
no other fidelities will occur.)

%%%% *NOT* /verification/
Many authors use the term (quantum state) verification for this or related tasks.  We want to save that word for
the motivation of our approach through the \emph{verification} (in the sense of ``testing'') of quantum computer
systems (cf.~\S\ref{ssec:intro:moti}; see Table~\ref{tab:vocabulary} for a dictionary).

%%%% Certification in related work
This definition QSC is that of Huang et al.~\cite{Huang-Preskill-Soleimanifar:certif-all:2024}.  It is
different from works where the only ``good'' state is the target state itself, or where a surviving state is
to be produced whose fidelity can be trusted to be high; see the discussion of related work
in~\S\ref{ssec:intro:rel-work}.

%%%% Near-term focus
As our method is motivated by the practical testing of near-term noisy quantum computer systems, we are
interested in a small number of qubits, e.g., $n \ll 1000$, and we assume that the time expended by repeated
shots with identical quantum circuits is negligible in comparison to the time cost of switching between
circuits, even if the modification of the circuit is only in the bases of the final single qubit
measurements.

%%%% Non-linearity as design goal
A design goal of the method we are proposing was to use measurement outcomes in a non-linear way, as this
allows the testing method --- at least in theory --- to catch some faults that may be missed by linear
averaging methods; see~\S\ref{sssec:intro:moti:nonlin} for details.

%%%% Why stabilizer states
We chose to work with stabilizer states as their certification is considerably easier, being amenable to simple
methods like direct fidelity estimation (DFE) \cite{Flammia-Liu:DirFidEest:2011}.  Even though Huang et
al.~\cite{Huang-Preskill-Soleimanifar:certif-all:2024} allow us to certify almost every state efficiently,
stabilizer state certification remains advantageous.  Indeed, Leone et
al.~\cite{Leone-Oliviero-Hamma:nonstabness-DFE:2022} proved that a certain measure of closeness to being a
stabilizer state determines the required repetition count for DFE.  Moreover, due to the ease of describing and
simulating Clifford circuits, in the context of testing, collections of test-circuits and state-descriptions can be
made available conveniently.
The initial goal of our research was to propose a practical method for testing quantum computer systems,
which implied that the whole testbed should be run quickly.  The long list of improvements in the state of
the art on certification-related problems for stabilizer states gives rise to the hope that expeditious
testing might be possible based on these advancements.

%%%% Contribution over most similar State Of The Art
The method we propose can be thought of as a modification of a method of Somma et
al.~\cite{Somma-Chiaverini-Berkeland:LB-fid-entangled:2006}, which takes a fixed basis\footnote{I.e., minimal
  generating set; we take the point of view that the stabilizer group is a vector space over~$\ZZ_2$, the field
  with~2 elements.} %
of the stabilizer group\footnote{In this paper, the term ``stabilizer group'' only appears as the stabilizer group
  of some stabilizer state, i.e., we are talking about groups of order $2^n$ of Hermitian $n$-qubit Pauli
  operators.} %
and computes an affine combination of the~$n$ estimates.
The contribution of this paper in terms of the process is (1) to choose the basis uniformly at random from
all bases of the stabilizer state, and (2) take the minimum of the estimates of the expectation values of
the observables in the chosen basis --- instead of the average --- to make the accept/reject decision.

%%%% Contribution in terms of theory
We analyze the probability that the method mistakenly accepts a bad state (i.e., one that has low fidelity
$\le 1-\eps$ with the target state) of rejects a good state (i.e., one that has high fidelity $\ge 1-\delta$ with
the target state).  As the method is randomized, there are two sources where the mistake could originate: Some of
the estimates of the expectation values could be off, or the choice of the basis might be bad.
The analysis of the former can be done using the usual tools in the usual way.  We refer to the latter source of
mistakes as ``intrinsic''.  Its analysis requires novel mathematical methodology.  The intrinsic mistake
probability depends, roughly\footnote{The truth is
  $\omega\delta(1 + \gamma_\good) = \beta = (1 - \gamma_\bad)\alpha\eps$, where the $\gamma_*$'s take care of
  statistical noise in the measurement results; $\beta$ is the threshold for deciding between accept and reject.},
on the quantities $\alpha < 1 < \omega$ in the equation $\alpha\eps = \omega\delta$: We can rule out intrinsic
false negatives deterministically at the cost of $\omega=2$; for intrinsic false positives we prove upper
($O((\alpha+1)/2)^n)$) and lower ($2^{-\Omega(n/\alpha)}$) bounds.  Yes --- the intrinsic false-positive
probability is\footnote{A useless observation in view of our stated target of $n\ll 1000$, which explains why the
  author has made no attempt to try to do the same with the intrinsic false negative probability.}
$o_{n\to\infty}(1)$ for fixed~$\alpha<1$.

While for a fixed basis, our bound involving the minimum of the expectation values is implied by Somma et al.'s
bound~\cite{Somma-Chiaverini-Berkeland:LB-fid-entangled:2006}, we find that taking the minimum is amenable to
amplification through randomization: For a fixed basis, in the worst-case over all bad states both the sum and the
minimum of the expectation values can be as low as $1-O(\eps/n)$ --- i.e., $\alpha = O(1/n)$ is needed, but
randomization ``amplifies'' the minimum to $1-\Omega(\eps)$ with overwhelming probability.

%%%% Remainder of this paper
\paragraph*{This extended abstract is organized as follows.}  %
The remainder of the introduction section is devoted to the motivation of our
approach~\S\ref{ssec:intro:moti}, the discussion of related work~\S\ref{ssec:intro:rel-work}, and a critical
assessment of the practical utility of the proposed method~\S\ref{ssec:intro:critic}.
Section~\ref{sec:BMoM}, we review DFE-based QSC, ``DFE-C'', as a starting point, and then present our method, BMoM
(``Basis Min-of-Means'').

In the full paper, in Section~\ref{FULL::sec:techout} we give a technical overview of the analysis of BMoM: It
contains the results that we have in full rigor, but in simplified form and without proofs; the full proofs are in
Section~\ref{FULL::sec:proofs}, and some particularly easy math was banished to the appendix.

\subsection{Motivation}\label{ssec:intro:moti}
%%%% This paper is motivated by:
This paper is motivated by the need to ``debug'' near-term, noisy quantum computer systems, which involve a
complex interplay of classical and quantum hardware devices, software, firmware across various subsystems.

%%%% *Faults* → automated *testing*
The assumption is that there are likely to be \textit{faults} (= ``bugs'', caused by human error) in the
design and/or its implementation of such a system, and that there is a need for automated testing.

%%%% *Verification*
This paper approaches these challenges in a manner analogous to design \emph{verification} in the classical
computing domain, except that, for some architectures, quantum computer systems are more flexibly
reconfigurable than classical chips, as a considerable part of the system is implemented in software.

%%%% *In Situ* (+ reconfig)
Design verification in the classical domain typically relies on simulations, which can already be expensive
even for classical systems.  For quantum computer systems, the cost of simulation may be prohibitive.  If a
fault involves interactions spanning more than just a few qubits, simulating the high-dimensional quantum
system quickly becomes infeasible.  As a result, certain aspects of verifying quantum computer system design
and implementation may need to occur \emph{in situ}, on the actual quantum device itself.  In these cases,
failures can only be observed in the quantum mechanical sense --- through measurements.
In the case of architectures where quantum computer systems are rapidly and inexpensively reconfigurable, in
situ verification is feasible, as it does not incur the penalties that render in situ testing prohibitively
expensive for classical chips.

%%%% Testing; *effectiveness & efficiency*
In this paper, the point of view is that verification through in-situ testing in the quantum case will
employ work similarly how testing works in the practice of classical software such as compilers: There is a
testbed of datasets with methods to verify correct outcome, concretely, in quantum, a testbed of quantum
circuits together with efficient quantum state certification.
This places a heavy burden on the need for quantum state verification procedures that are both efficient
(i.e., the whole testbed should complete in a short amount of time) and effective, achieving extremely low
rates of false positives and false negatives, for reliable quantum computer system verification.

%%%% *𝜺,𝜹* and rel to effectiveness, efficiency
In terms of the $\eps,\delta$ parameters for quantum state certification, a fraction $\eps/\delta$ that is too big
will generally jeopardize effectiveness, one that is too small might adversely affect efficiency; hence we require
that $\eps/\delta$ is (near) a small\footnote{A typical choice in DFE-C would be $\eps=2\delta$; the corresponding
  choice the method in this paper would be $\eps=8\delta$, if the number of qubits in the high double digits (the
  gap gets smaller with more qubits).}  constant.

%%%% Testbed: *Clifford* circuits
As explained above, in this paper we focus on testing through preparing stabilizer states, so the testbed would
consist of a collection of Clifford circuits.

%%%% *Effectiveness*
In terms of effectiveness, methods for QSC are, of course, subject to stochastic noise in the measurement
outcomes, but they may also involve intrinsic variability due to the protocol itself employing randomness.
In this paper, we adopt the perspective that, for QSC to be considered effective, the effect of both
measurement noise and intrinsic randomness must be rigorously bounded.

%%%% *Efficiency* & cost model
In terms of efficiency we require single qubit measurements, as is customary in the state of the art.  We assume
that the cost of repeatedly executing a quantum circuit without changes --- ``shots'' --- is negligible compared to
changing the measurement settings (which qubit is measured in which basis) of the circuit.  This is in line with
some approaches to QSC, e.g., \cite{ Tiurev-Sorensen:fid-meas-cluster-min-effort:2021,
  Aktar-Bärtschi-Badawy-Eidenbenz:experiment-bd-fid:2022,
  Barbera-Rodriguez-Navarro-Zambrano:groups-enhance-DFE:2025}.
Finally, we are interested in the few-qubit regime, say, $n \ll 1000$, so that a linear scaling with~$n$ is
acceptable.
We note that simply using vanilla Direct Fidelity Estimation~\cite{Flammia-Liu:DirFidEest:2011} for
certifying a known stabilizer state requires\footnote{Note that $\delta = \Omega(\eps)$ allows the use of
  the multiplicative Chernoff bound.} $O(\ln(1/p)/\eps)$ measurement settings --- the same as the total
number of shots.

\subsubsection{Non/linear certificate functions}\label{sssec:intro:moti:nonlin}
%%%% Deaf: *Certificate*
The word ``certification'' in QSC suggests that a ``certificate'' is produced: In a typical (non-adaptive)
protocol for quantum state certification, a function is employed that maps sequences of binary measurement
outcomes to a numerical values.  The accept/reject decision is then made based on the value of the function,
and the collection of measurement results would be the certificate.

%%%% Certificate *function*
The \emph{certificate function} could be, e.g., a weighted average.  For effective bug-catching, it may
matter whether it's linear or not, as we will now elaborate.

%%%% *Non/linear* certificate function
Whether a fault causes a failure --- i.e., an observable incorrect behavior --- can depend on the run-time
context in complex ways.  In quantum mechanical terms, a sate could be prepared incorrectly as the effect of
a fault, and over several iterations, the fault could manifest in different effects on the state.  The
deeper the loop, the more likely these pseudo-non-deterministic effects are.  This can be a problem for
verification if the effects on the fidelity cancel out: the empirical mixed state that is prepared over
several system runs (i.e., the mean over the runs of the state prepared in the run) lies within the fidelity
bounds that must be accepted as they could be caused by quantum error.

One possibility is to make a deeper picture of the state, not just its fidelity, e.g., using tomography.  In any
case, though, any protocol that makes an accept/reject decision (whether it targets fidelity or not) which
processes the measurement results in an affine way (i.e., the certification function commutes with linear
combinations whose coefficients sum to 1) cannot distinguish between quantum error and effects of faults.  One
possibility is to take a selection of observables and iterate over them in an outer loop, and for each of them
estimate the expectation value in an inner loop.  While the inner loop would typically be linear, the outer loop
might not be.  E.g., in state tomography, a norm (non-linear!) of the estimated tomography coefficients with the
target coefficient might be computed.

In the method we propose the outer loop takes the minimum over the chosen observables of the expectation
values: The certificate function is the minimum of means of measurement results.

\begin{figure*}[tp]
  \centering
  \begin{subfigure}{0.48\textwidth}
    \centering
    \includegraphics[width=\textwidth]{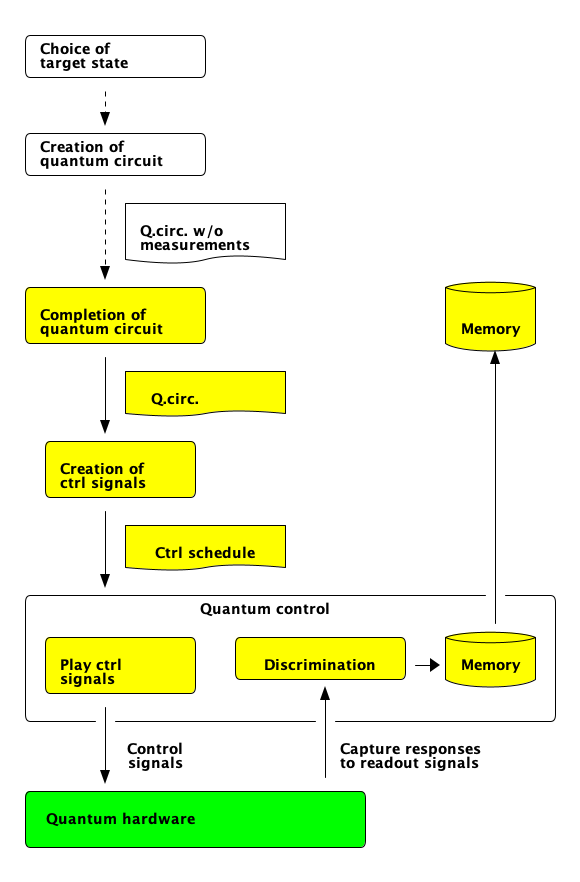}
    \caption{NISQ-era Quantum Computer System Diagram}\label{fig:nisqc_system_testing:long}
  \end{subfigure}
  \hfill
  \begin{subfigure}{0.48\textwidth}
    \centering
    \includegraphics[width=\textwidth]{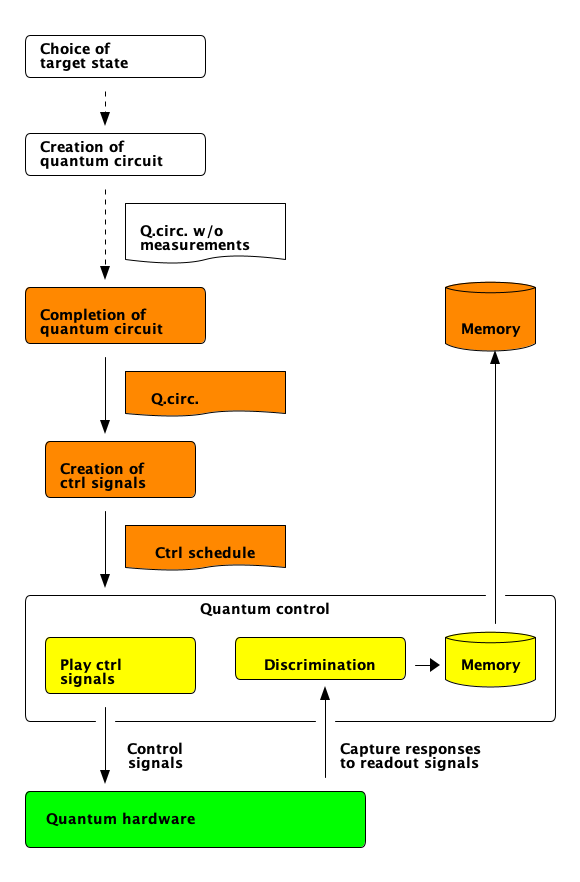}
    \caption{NISQ-era Quantum Computer System Diagram with shallow loop}\label{fig:nisqc_system_testing:short}
  \end{subfigure}
  \mycaption{NISQ Computer-System Testing.}{%
    This is a simplified system diagram of an example NISQ-era quantum computer system.
    (a) Unavoidable quantum error happens in the quantum hardware (green).  Faults can be present in any of the
    yellow-colored components (boxes).
    Certification functions that are affine in the measurement results (such as averages) cannot distinguish
    between the effect of faults and the effect of quantum error \emph{unless} the effect of the fault throws the
    empirical fidelity (fidelity of the mixed state that is the mean of the states generated in the course of the
    protocol) off.
    (b) Certification functions that process estimates of expectation values for different observables in a
    non-linear way have at least a theoretical chance of distinguishing between the effect of quantum error and the
    effect of faults: in the case that the fault throws only some of the expectation values off.  Hence faults in
    the orange boxes have at least a theoretical possibility of being caught even if they cancel out over the
    totality of the observables that are processed.
  }\label{fig:nisqc_system_testing}
\end{figure*}

%%%% Ref *system diagram* figure for non/linear
We hypothesize (but defer to future testing) that this non-linearity can catch faults that linear certificate
functions miss.  Fig.~\ref{fig:nisqc_system_testing} illustrates which faults might in theory be caught by
non-linear processing of expectation values of observables.

\begin{table}[tp]
  \centering
  \begin{tabular}{lp{11cm}}
    \hline
    \textbf{Term}     & \textbf{Explanation} \\
    \hline
    \it Mistake       & Certification method gives wrong result, i.e., \\
    {}                & \quad false positive or false negative\\[1ex]
    \it Noise         & Statistical deviations of estimates from expectations \\
    {}                & \quad caused by randomness of measurement outcomes\\[1ex]
    \it Quantum error & Caused by small defects in quantum hardware, interaction\\
    {}                & \quad with bath, etc.\\[1ex]
    \it Fault         & $=$ ``Bug'', flaw in the hard- or software caused by human error\\
    {}                & \quad in design or implementation\\[1ex]
    \it Failure       & Observable incorrect behavior caused by a fault\\%[1ex]
    \hline
  \end{tabular}
  \mycaption{Encyclopedia of wrong.}{%
    ``Mistake'' refers to the event that a certification method produces a wrong result, i.e., a false positive
    (accepting a bad state) or a false negative (rejecting a good state).  %
    ``Noise'' is statistical noise in measurement results (via Born rule).  %
    ``Quantum error'' arises from imperfections in the hardware or unavoidable coupling with the environment.  %
    ``Fault'' and ``failure'' follow classical verification usage: A ``fault'' is a flaw in the hard- or software
    caused by a human error, leading to incorrect behavior under certain conditions, while ``failure'' is the
    observable incorrect behavior of the system when a fault is triggered during execution (e.g., incorrect output
    in an FPGA due to a timing violation).
    ``Bug'' is an informal term for a fault.
    \\
    We emphasize that in this paper, ``faults'' are classical bugs caused by human error, not impurities arising in
    the fabrication etc of the quantum device.
    \\
    As an example of the terminology, we note that the BMoM method has two sources where mistakes may come from:
    The statistical noise in the estimation of expectation values and the random choice of the basis, which may
    lead to the expectation values themselves causing the mistake.
  }\label{tab:vocabulary}
\end{table}

\subsection{Related work}\label{ssec:intro:rel-work}
%
% Related Methodology or Analysis
%
As mentioned above, the concept of state certification is that of Huang et
al.~\cite{Huang-Preskill-Soleimanifar:certif-all:2024}, whom we also follow in taking the infidelity (i.e., one
minus fidelity) upper bound for good states within a constant of the infidelity lower bound for bad states.  In
this realm, multiplicative Chernoff bounds give us that an average of $O(\log(1/p)/\eps)$ copies of a bounded
random variable allows to distinguish between the good and bad cases with probability $1-p$, where~$\eps$ is the
infidelity bound (either of them).

This concept is different from works where the only ``good'' state is the target state itself (or, in our logic,
the infidelity upper bound for good states is little-o of the infidelity lower bound for bad states), e.g.,
\cite{Pallister-Linden-Montanaro:opt-veri-ent-loc:2017, Takeuchi-Morimae:verif-manyqubit:2017,
  Markham-Krause:simple-cert-graphstates:2018, Zhu-Hayashi:general-verif-advers:2019,
  Dangniam-Han-Zhu:optim-verif-stab:2020}

This concept is also different from those where a surviving state is to be produced whose fidelity can be trusted
to be high, under various assumptions~\cite{Takeuchi-Morimae:verif-manyqubit:2017,
  Markham-Krause:simple-cert-graphstates:2018, Zhu-Hayashi:general-verif-advers:2019,
  Dangniam-Han-Zhu:optim-verif-stab:2020}.

The taxonomy of Gherghiu et al.'s 2017 overview~\cite{Gheorghiu-Kapourniotis-Kashefi:verif-QC-overview:2017} does
not directly encompass our setting, but our approach aligns most closely with the sub-universal protocols and
experimental fault-tolerance techniques discussed in their Sections 5.1 and~5.2.

\mypar
%
% Emphasize need for few observables
%
Next to the Huang et al.'s paper and Flammia \& Liu's~\cite{Flammia-Liu:DirFidEest:2011} and da Silva et al.'s
\cite{daSilva-LandonCardinal-Poulin:pract-char-qdev-w/o-tomo:2011} original treatments of Direct Fidelity
Estimation --- we review the application to state certification in \S\ref{ssec:BMoM:dfe} ---, the primary
inspirations of this paper include the seminal treatment of Somma et
al.~\cite{Somma-Chiaverini-Berkeland:LB-fid-entangled:2006} for stabilizer states.

Somma et al.\ prove a lower bound on the fidelity of an unknown state~$\varrho$ with a known $n$-qubit stabilizer
state~$\psi$ based on an affine combination of the observables in a fixed basis of the stabilizer group.  Expressed
in terms of infidelity, their bound reads:
\begin{equation}\label{eq:somma-infidelity-UB}
  \eps \le \tfrac{n}{2} (1 - \mu)
\end{equation}
where $\mu$ is the mean of expectation values of the~$n$ observables in the fixed basis of the stabilizer group.
Somma et al.'s upper bound on infidelity implies our ``Deterministic Basis Bound''
(Proposition~\ref{FULL::prop:deterministic-basis-bound}, \S\ref{FULL::sssec:techout:falsepos:determini} of the full
paper).  While our bound is weaker, as it simply replaces the mean by the minimum, it lends itself to
\emph{amplification through randomization}: Taking a random basis instead of any old fixed basis reduces the factor
in $\eps \le \tfrac{n}{2}(1 - \nu)$ (where $\nu$ is the minimum) from $O(n)$ to $O(1)$.

Somma et al.'s result was used by Aktar et al.~\cite{Aktar-Bärtschi-Badawy-Eidenbenz:experiment-bd-fid:2022} to
experimentally lower-bound fidelities of entangled states prepared on near term noisy quantum computing devices.
Moreover, it was rediscovered in another form in~\cite{Pallister-Linden-Montanaro:opt-veri-ent-loc:2017}, and
generalized, e.g., \cite{Kalev-Kyrillidis-Linke:val+cert-stab:2018,
  Zhou-Zhao-Yuan-Ma:detect-multipart-entangle:2019}.

As Somma et al., both Pallister et al.~\cite{Pallister-Linden-Montanaro:opt-veri-ent-loc:2017} and Kalev et
al~.\cite{Kalev-Kyrillidis-Linke:val+cert-stab:2018} base their methods on measuring the observables in a single
basis of the stabilizer group.  This is similar to the present work, but we take a random basis.

\mypar
%
% Need for few observables
%
Somma et al.'s~\cite{Somma-Chiaverini-Berkeland:LB-fid-entangled:2006} work is further related to the present paper
in that they acknowledge the need to keep the number of different measurement settings low.  Although the necessity
for conducting only local measurements is commonly recognized, the specific challenge of reducing measurement
configurations is not as prominent.  With Somma et al., as mentioned above,
\cite{Pallister-Linden-Montanaro:opt-veri-ent-loc:2017,Kalev-Kyrillidis-Linke:val+cert-stab:2018} use~$n$
measurement settings.  Zhou et al.~\cite{Zhou-Zhao-Yuan-Ma:detect-multipart-entangle:2019}, Tiurev \&
S\o{}rensen~\cite{Tiurev-Sorensen:fid-meas-cluster-min-effort:2021}, and, notably, the experimental
paper~\cite{Aktar-Bärtschi-Badawy-Eidenbenz:experiment-bd-fid:2022} all go below~$n$.  Indeed, the concept of
combining observables that can be simultaneously locally diagonalized into a single measurement was already worked
out by Somma et al.  For graph states, \cite{Zhou-Zhao-Yuan-Ma:detect-multipart-entangle:2019,
  Tiurev-Sorensen:fid-meas-cluster-min-effort:2021} combine the stabilizer generators within vertex-coloring
classes.  These works prefer the cost of quadratic in~$n$ worst-case scaling of the total number of shots, at the
benefit of reducing the number of observables to $o(n)$ or even
\cite{Aktar-Bärtschi-Badawy-Eidenbenz:experiment-bd-fid:2022,
  Barbera-Rodriguez-Navarro-Zambrano:groups-enhance-DFE:2025} $O(1)$.

While some of these works argue that the savings in variance reduce the shot count, it is noteworthy that rigorous
bounds are not available for stabilizer states (but see Barber\`a-Rodr\'iguez et
al.~\cite{Barbera-Rodriguez-Navarro-Zambrano:groups-enhance-DFE:2025} for other types of states).  In the present
paper, we take the standpoint that, in order to have reliable tests --- i.e., a false positive or false negative
cannot be caused by a type of fault or a type of good state that was under-represented in simulations ---
mathematically rigorous worst-case bounds must be used.

\mypar
%
% Non-linearity of measurement processing
%
In terms of the of certificate functions, in the setup where good states can have non-negligible infidelity with
the target state, mostly affine functions are used.  This includes Huang et
al.~\cite{Huang-Preskill-Soleimanifar:certif-all:2024} as well as Somma et
al.~\cite{Somma-Chiaverini-Berkeland:LB-fid-entangled:2006} and directly related methods
\cite{Zhou-Zhao-Yuan-Ma:detect-multipart-entangle:2019, Aktar-Bärtschi-Badawy-Eidenbenz:experiment-bd-fid:2022},
and also DFE and derivatives
\cite{daSilva-LandonCardinal-Poulin:pract-char-qdev-w/o-tomo:2011,Flammia-Liu:DirFidEest:2011,
  Tiurev-Sorensen:fid-meas-cluster-min-effort:2021}.  Non-affine certificate functions appear where the target
state is (up to phase and negligible infidelity $\delta\to 0$) the only good state: There, a reject answer is given
if a single (binary) measurement result deviates from that of the target stabilizer state; this makes the
certification function a non-linear polynomial in the measurement results.

A beautiful exception to the simplicity of certification functions is Kalev et al.'s
work\cite{Kalev-Kyrillidis-Linke:val+cert-stab:2018}: Through a semidefinite programming argument, they determine
the smallest fidelity that a state could have, subject to fixing its expectation values on the observables in a
fixed basis of the stabilizer group.  It would be extremely interesting to find a way to combine Kalev et al.'s
bound with our amplification through randomization concept.

\mypar
%
% Doesn't really fit anywhere
%
Derbyshire et al.~\cite{Derbyshire-Mezher-Kapourniotis-Kashefi:rnd-bench-stab-verif:2021} propose a
method combining randomized benchmarking with stabilizer-based verification, although their emphasis rests more on
gate-synthesis aspects than on fidelity bounds in the QSC sense.  Takeuchi et
al.~\cite{Takeuchi-Takahashi-Morimae-Tani:div-conq-verif:2021} develop a compelling ``divide-and-conquer''
framework for verifying shallow circuits under stringent hardware constraints on measurements, but their focus
is not directly related to the approach considered here.

\subsection{Critical evaluation}\label{ssec:intro:critic}
While the present paper offers a novel approach to quantum state certification, backed up by rigorous mathematical
analysis involving tools and results that might conceivably be of independent interest, the practical utility of
the proposed method --- measured on the goal of enabling effective and efficient tests of quantum computer systems
--- is somewhat questionable.  In this subsection, we discuss the merits of the present paper critically.

\subsubsection{Shortcomings of the theoretical analysis}\label{sssec:intro:critic:analysis}
No attempt has been made to improve the deterministic factor~$\omega=2$ for the gap on expectation bound for good
states to remove intrinsic false negatives: Amplification through randomization will likely work here, too,
reducing~$\omega$ to a number in $\lt[1,2\rt]$ with high probability (rather than deterministically) over the
randomly chosen basis.

Additionally, while we provide lower and upper bounds for the probability of an intrinsic false positive as a
function of the number of qubits~$n$ and the multiplicative gap~$\alpha<1$ between the good and bad state
fidelities, these bounds are quite far apart.

A related issue is that we are not presenting our best lower bound, as we were unable to find a closed-form
expression for it.  The bound is connected with open problems in enumerative combinatorics that involve non-square
matrices with full rank and limits on the number of non-zero entries.

Finally, we cannot yet back up with either (non-trivial) analysis or numerical simulations our hope expressed in
\S\ref{sssec:intro:moti:nonlin} above, that our non-linear certificate function will be effective in identifying
some faults that would have been missed with a linear approach.

\subsubsection{Disadvantages of the method}\label{sssec:intro:critic:method}
Known state certification methods --- such as, for concreteness, DFE-based QSC (DFE-C), have one source for
mistakes: The statistical noise in the estimation of the mean value.  BMoM has ``intrinsic'' mistakes as an
additional source.  Methods such as DFE-C require a gap between the fidelity of a good state and a bad state to
avoid mistakes based on statistical noise, but BMoM requires a bigger gap to also accommodate the intrinsic
mistakes with high probability.
This phenomenon is highlighted by the fact that for DFE-C, the only requirement is $\delta < \eps$ --- if blowing
up the total number of shots is acceptable.  But BMoM has harsher requirements, e.g., as presented here, it can
only work if $\omega\delta < \alpha\eps$.

It could be argued that BMoM sits between two seats: While it does reduce the number of required measurement
configurations from DFE-C, it reduces it only to the number of qubits~$n$.  But there are other methods that reduce
the number of different measurement settings to $O(1)$.  (These methods may have their own disadvantages, notably
they all seem to require $\eps = \Omega(n\delta)$.)

While the use of a minimum in the certificate functions does lend itself to amplification through randomization, it
has the disadvantage that each of the~$n$ expectation values need to be accurately estimated, leading to an
additional $\log n$ cost (union bound) in the total number of shots.  We note that the minimum could be estimated
with a smaller number of shots (e.g., using multi-armed bandits methods), but that would come at a cost of
increasing the number of measurement settings.

See Table~\ref{tab:dfe-M--vs--bmom-m} for an overview of repetition counts for outer loop (over the observables)
and inner loop (shots with same circuit).

\begin{table}[htp]
  \centering
  \begin{tabular}{lcc}
    \hline
    {}     & Outer Loop    & Inner Loop \\
    {}     & (observables) &  (shots)   \\
    \hline
    DFE-C  &    $M$        &   $1$      \\
    BMoM   &    $n$        &   $m$      \\
    \hline
  \end{tabular}
  \mycaption{Repetition parameters in DFE-C vs BMoM.}{%
    The letter~$n$ stands for the number of qubits.  By default, DFE runs a single (``outer'') loop over final
    measurements, repeating each measurement once (``inner loop'' has 1 iteration).  In contrast, BMoM selects as
    many final measurements as there are qubits --- under the near-term assumption that $n \ll M$, and repeats each
    measurement several times.  Increasing~$M$ or~$m$, resp., reduces the probability that stochastic noise
    introduced by quantum mechanical measurements leads to a mis-classification (a method ``mistake'', in our
    terminology).  As the interval available to accommodate stochastic noise is smaller in BMoM (it is reduced
    already though the need to reduce mistakes due to a bad choice of the~$n$ measurements), we will generally have
    $M \ll m$ --- and the $\log n$ cost of the union bound increases the inequality.}\label{tab:dfe-M--vs--bmom-m}
\end{table}

% ¤intro:critic:method:conclusion
In summary, BMoM is attractive only under two quantitative conditions:
\begin{enumerate}[label=C\arabic*)]
\item \textbf{Few qubits.}  The number of qubits~$n$ must remain in the small-scale regime ($n\ll 1000$) ---
  otherwise DFE-C's total shot count is near~$n$, and BMoM offers only disadvantages.
\item \textbf{Wide certification gap.}  The interval $[\delta,\eps]$ must satisfy $\eps > 2\delta$ --- in practice
  probably $\eps = 8\delta$ (depending on~$n$, getting better with bigger~$n$) --- so that the intrinsic mistake
  probability of BMoM can be pushed below the acceptable rate; in comparison, DFE-C typically works with
  $\eps = 2\delta$.
\end{enumerate}

But meeting C1--C2 is only half the story.  The intended speed-up comes from re-using an \emph{identical} circuit
in an inner loop while changing only the final measurement basis in the outer loop.  This may pays off in one of
two scenarios:
\begin{itemize}
\item[\emph{(i)}] \emph{Latency-dominated devices.}  On current-generation hardware, recompiling control waveforms
  --- even just for a different Pauli basis --- can be considerably slower than firing additional shots with a
  frozen schedule.  If that ratio persists, BMoM reduces wall-clock time despite its larger shot count.
\item[\emph{(ii)}] \emph{Hierarchical debugging.}  When the goal is to expose faults in the higher-level components
  (orange in Fig.~\ref{fig:nisqc_system_testing}) a ``non-linear variant'' of DFE graph remains a natural strategy
  --- and the extra classical runtime would be acceptable.  But at this time, this ability of BMoM to expose these
  faults is merely a hypothesis.
\end{itemize}

In any case, both advantages may evaporate on future NISQ machines: Error-mitigation frameworks such as randomized
compilation continuously mutate the circuit, and tighter classical-quantum integration will move changes in the
measurement configuration into the yellow area of Fig.~\ref{fig:nisqc_system_testing}.

\section{DFE-C and BMoM}\label{sec:BMoM}
\subsection{DFE-based certification of a stabilizer state}\label{ssec:BMoM:dfe}
Before we present the Basis-Min-of-Means method, for comparison, we review state certification based on DFE in
the case of stabilizer states in Algorithm~\ref{alg:dfe}.

\begin{algorithm}
  \caption{DFE-C$(n, M, \beta, C)$}\label{alg:dfe}
  \begin{algorithmic}[1]
    \Require $\bullet$~\ $n$ (number of qubits), circuit template~$C$; parameters $M$, $\beta$. \\
    $\bullet$~\ Ability to sample indep.~uar from the stabilizers of the state prepared by~$C$.
    \For{$M$ times}%
      \State Sample a stabilizer~$x$%
      \State Extend $C$ with the stabilizer measurement $\to C'$                       \label{alg:dfe:prep}%
      \State Run the circuit~$C'$ once, collecting the $\pm1$ measurement result%
    \EndFor
    \State Compute the mean of the measurement results                                 \label{alg:dfe:mean}%
    \If{the mean $\ge$ $1-\beta$} \texttt{ACCEPT} \textbf{else} \texttt{REJECT}.
    \EndIf
  \end{algorithmic}
\end{algorithm}

The Algorithm gets as input a number~$n$ of qubits, a discrimination threshold~$\beta$, a number~$M$ of repetitions
of the loop, and a quantum circuit template~$C$.  By ``template'', we mean a circuit that prepares a state on
the~$n$ qubits.  It may contain initializations of the~$n$ qubits and unitary
operations; %%, #FeedForward and internal measurements whose outcomes are feed-forwarded.
we require that the state prepared by~$C$, if run ideally (i.e., without faults or quantum error), is a stabilizer
state.

Finally, the algorithm also requires the ability to sample repeatedly, independently uniformly at random from the
stabilizer group of the state prepared by~$C$; we assume that this capacity is given as an oracle.

\myparsmall%
To broaden the applicability of the method, we might allow that~$C$ operates on more than~$n$ qubits, but prepares
a product state on two registers, one of them has~$n$ qubits and is in a stabilizer state at the end of~$C$ --- a
change irrelevant to the methods and their analysis.
%
% Moreover, we purposefully neglect to allow the prepared state to depend on the outcomes of the mid-circuit
% feed-forward measurements, as that would clutter up the presentation of the methods.  % #FeedForward

\myparsmall%
Assuming $\eps,\delta,\beta$ are constants independent of~$n$ and $\beta \in \ltoI\delta,\eps\rtoI$, it is straight
forward that DFE-C is correct up to statistical noise, which decreases as $O(1/M)$.  This is owed to the fact that
the (operator-valued) expectation of an operator picked uar from a stabilizer group is equal to the projector onto
the stabilized space, along with the usual (multiplicative) Chernoff bounds.

The quantities $M,\beta$ can be derived from the fidelity bound quantities $\delta,\eps$ and a upper bound~$p$ on
an acceptable probability of a mistake in the result of a certification; we explain how.
Suppose that the circuit~$C$, if run ideally, produces a stabilizer state~$\psi_0$, and if it is run without faults
but with quantum error produces a mixed state~$\varrho_0$.
In the context of testing, the state~$\varrho_0$ is often not known and its properties can only be approximately
estimated based on simulation with error models~\cite{Huang-...-Kockum-Tancredi:proc-tomog-digi-twin:2025}.  As
outlined in the introduction, in this paper, we assume that two fidelity bounds are available.  Assuming
$\varrho_0$ is unknown, we define the following in general:
\begin{itemize}
\item Any state~$\varrho$ with $\bra{\psi_0}\varrho\ket{\psi_0} \le 1-\eps$ is considered to be affected by a
  fault, and must be rejected: By definition, it is a \textit{bad state};
\item Any output state~$\varrho$ with $1-\delta \le \ket{\psi_0}\varrho\bra{\psi_0}$, is considered not to be
  affected by a fault, and must be accepted: By definition, it is a \textit{good state.}
\end{itemize}
We give an example of how $\beta,M$ could be determined.  Given any value of $\beta \in \ltoI\delta,\eps\rtoI$ and
probability target~$p_\good$, one uses Chernoff's theorem to find the number of repetitions $M_\good$ that are
required such that if Alg.~\ref{alg:dfe} prepares (on average) a state of fidelity $1-\delta$ in
Step~\ref{alg:dfe:prep} --- i.e., the ``worst'' good state --- the Chernoff upper-tail bound for the probability
that the mean in Step~\ref{alg:dfe:mean} is less than $1-\beta$ is just below~$p_\good$.  When running
Alg.~\ref{alg:dfe} with this~$\beta$ and $M:=M_\good$ we can be sure the probability of a false negative is below
$p_\good$.
We proceed symmetrically given $p_\bad$, and the number of shots $M_\bad$ for the case when the ``best'' bad state
--- with fidelity $1-\eps$ --- is prepared.  Combining the two bounds, given an upper bound~$p$ on the total
probability that the certification method makes a mistake, to minimize the number of repetitions, it is prudent to
have $M_\good = M_\bad$, which we can achieve by making $p_\good,p_\bad$ variable and choosing~$\beta$ subject to
$M_\good = M_\bad$ and $p=p_\good +p_\bad$.  This results in gaps $(1+\gamma_\good)\delta = \beta$ and
$(1-\gamma_\bad)\eps = \beta$, where $0 < \gamma_\good$, and $0<\gamma_\bad < 1$.

There are other methods to finding $\beta,M$.  The point is that there are (at least) two ``implicit'' parameters
$\gamma_\good$ and $\gamma_\bad$ that affect the performance of the method.  For DFE-C, without a bound on~$M$,
only $\delta<\eps$ is required for this to work, and if $\eps = O(\delta)$, it results in repetition numbers
$M = O(\log(\nfrac1p)/\delta)$.

\subsection{The Basis-Min-of-Means approach}\label{ssec:BMoM:bmom}
For BMoM, there will be more implicit parameters, and, unfortunately, harsher constraints on $\delta,\eps$.  Let's
start by discussing the method itself.
\subsubsection{The method}\label{sssec:BMoM:bmom:method}
Algorithm~\ref{alg:bmom} shows the Basis-Min-of-Means approach.  As for DFE-C, we present it first with the
abstract parameters $m,\beta$, and discuss below how these parameters should be chosen depending on the fidelity
bounds and mistake-probability targets.
% (To preempt questions about $m$ vs.~$M$, we refer to Table~\ref{tab:dfe-M--vs--bmom-m}.)

\begin{algorithm}
  \caption{Basis-Min-of-Means$(n, m, \beta, C)$}\label{alg:bmom}
  \begin{algorithmic}[1]
    \Require $\bullet$~\ $n$ (number of qubits), circuit template~$C$; parameters $m$, $\beta$. \\
    $\bullet$~\ Ability to sample indep.~uar from the stabilizers of the state prepared by~$C$.
    \State Sample a random basis of the stabilizer group.%
    \For{each stabilizer in the basis}%
      \State Extend $C$ with the stabilizer measurement $\to C'$%
      \State Run the circuit~$C'$ $m$ times, collecting the  $\pm1$ measurement results%
      \State Compute the mean of the measurement results%
    \EndFor%
    \If{minimum of the means $\ge$ $1-\beta$} \texttt{ACCEPT} \textbf{else} \texttt{REJECT}.
    \EndIf
  \end{algorithmic}
\end{algorithm}

Let us go through Algorithm~\ref{alg:bmom} in detail.  Just as DFE-C, the algorithm gets as input a number~$n$ of
qubits, a discrimination threshold~$\beta$, and a quantum circuit template~$C$.  Different from DFE-C, BMoM gets
the number~$m$ of repetitions of the \emph{inner loop,} i.e., \emph{shots with the same circuit}, $C'$.

BMoM also needs a way to sample uniformly at random from the bases of the stabilizer group of the state.  Uniformly
sampling a basis can be realized by simply sampling~$n$ stabilizer group elements independently uniformly at
random, and repeating until the set of stabilizers forms a basis.  Indeed, it can be readily checked that the
number of repetitions follows a geometric distribution whose parameter decreases towards
$\phi(\nfrac12) \approx 0.2885$ as $n\to\infty$, where~$\phi$ is the Euler
function\footnote{\url{https://en.wikipedia.org/wiki/Euler_function}}, i.e., the expected number of repetitions
increases towards $1/\phi(\nfrac12) \approx 3.5$.

After randomly selecting a basis of the stabilizer group, the algorithm goes through the stabilizers one by one: a
circuit~$C'$ is formed by appending instructions to measure the stabilizer observable to the circuit template.
Typically this will happen by measuring all~$n$ qubits individually in suitable bases.  The circuit~$C'$ is then
executed~$m$ times, and the~$m$ $\pm1$-values are averaged to form an estimate of the expectation value of the
stabilizer observable.

If at least one of the~$n$ averages is below the threshold $1-\beta$, the method rejects; if on the other hand all
the estimates are greater than or equal to $1-\beta$, the method accepts.

\subsubsection{The parameters}\label{sssec:BMoM:bmom:param}
We now explain how to obtain the parameters $\beta$ and~$m$.  Where for DFE-C we had two implicit parameters
$\gamma_\good$ and $\gamma_\bad$ that required deciding on in order to find $\beta,M$, for BMoM, have one more,
$\alpha$, see Table~\ref{tab:bmom-params}; one other possible parameter (without letter) is fixed to~2.

\begin{table}[tp]
  \centering
  \begin{tabular}{ll}
    \hline\hline\\[-2ex]
    \multicolumn{2}{l}{\underline{Given parameters:}}\\[.5ex]
    $\delta \in \ltoI 0     ,1\rtoI$ & Fidelity bound for good states \\
    $\eps   \in \ltoI\delta,1\rtI$   & Fidelity bound for bad states  \\
    $p \in \ltoI 0,1\rtI$            & Upper bound to probability of mistake \\
    \hline\hline\\[-2ex]
    \multicolumn{2}{l}{\underline{Implicit parameters:}}\\[.5ex]
    $\gamma_\good \in \ltoI 0,\infty\rtoI$ & Gap from expectation to accommodate statistical noise towards good state \\
    $\gamma_\bad  \in \ltoI 0,1     \rtI$  & Gap from expectation to accommodate statistical noise towards bad state \\
    $\alpha \in \ltoI 0,1\rtoI$            & Gap (multiplicative) on expectation bound for bad states to remove \\
    {}                                     & intrinsic false positives \\
    \hline\hline\\[-2ex]
    \multicolumn{2}{l}{\underline{Possible additional implicit parameter:}}\\[.5ex]
    $\omega$ ($:=2$) $\in \ltoI 1, \infty \rtoI$  & Gap (multiplicative) on expectation bound for good states to \\
    {}                                            & remove intrinsic false negatives \\
    \hline\hline
  \end{tabular}%
  \mycaption{Parameters of DFE-C and BMoM.}{The threshold parameter~$\beta$ along with the number of repetitions of
    the outer loop~$M$ (DFE-C) or the inner loop $m$ (BMoM) are determined from the given fidelity parameters
    $0<\delta<\eps$ and an bound on an acceptable probability of a mistake of the certification method.  This is
    done by deriving values for the ``implicit'' parameters.  Both DFE-C and BMoM require space to accommodate
    statistical noise --- $\gamma_\good$, $\gamma_\bad$ --- but only BMoM requires space to accommodate
    ``intrinsic'' mistakes: $\alpha$ and $\omega$.  In this paper we set $\omega := 2$.  For DFE-C, we have
    $\delta\cdot(1+\gamma_\good) = \beta = (1-\gamma_\bad)\cdot\eps$; for BMoM we have
    $\omega\delta\cdot(1+\gamma_\good) = \beta = (1-\gamma_\bad)\cdot\alpha\eps$.
  }\label{tab:bmom-params}
\end{table}

\begin{figure}[tp]
    \centering
    \includegraphics[width=0.8\textwidth]{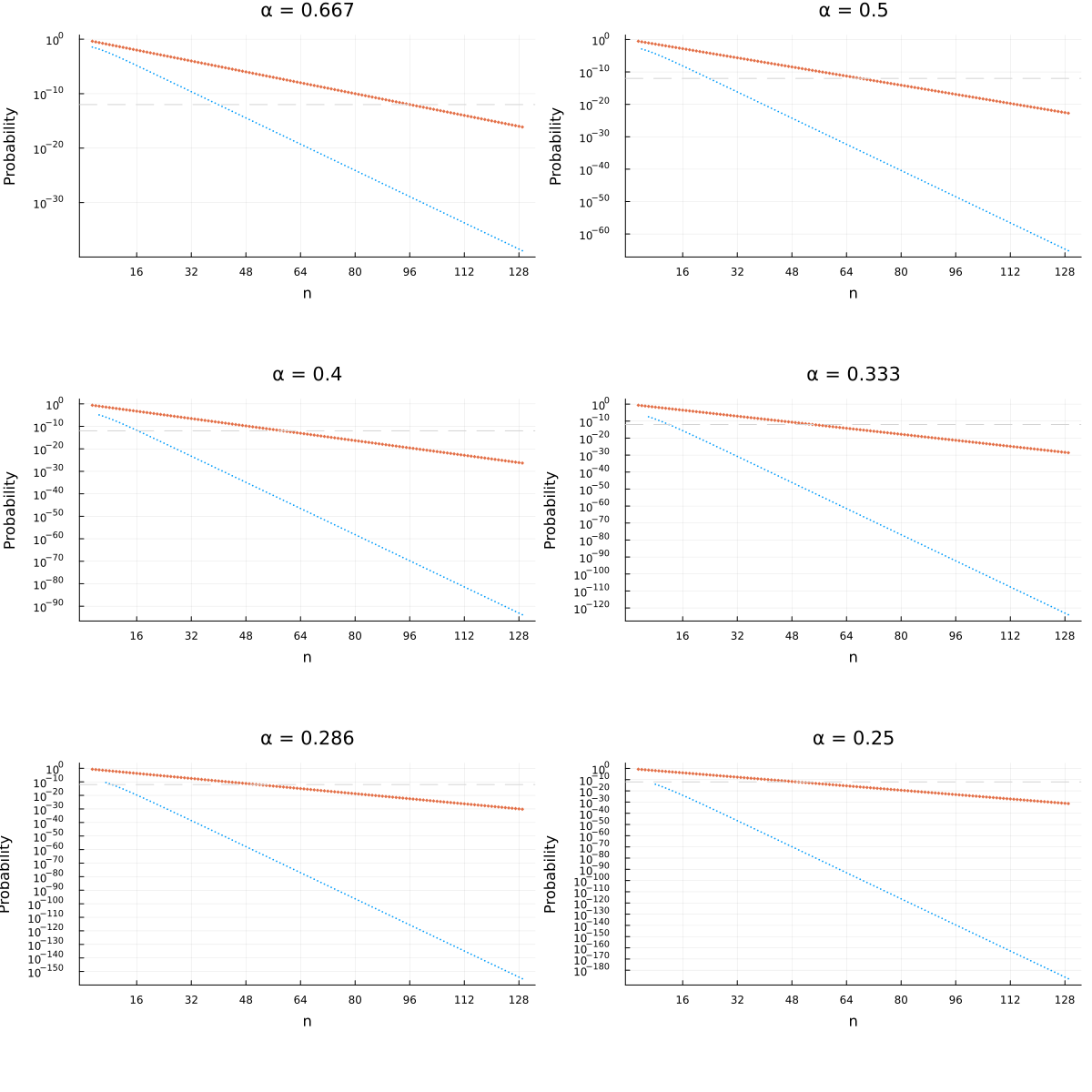}
    \mycaption{Intrinsic false-positive probabilities: Lower and upper bounds.}{%
      The plots show the intrinsic false-positive probability lower and upper bounds in the form in which they are
      proven in~\S\ref{FULL::ssec:pf:α:probab} of the full paper , see Propositions
      \ref{FULL::prop:pf:α:probab:thm-UB} and~\ref{FULL::prop:pf:α:probab:thm-LB} there.  The probability
      $10^{-12}$ is marked with a dashed horizontal line, as an example of an acceptable rate of intrinsic false
      positives.
      It could be concluded that for more than 64 qubits, $\alpha = 1/2$ would be acceptable; that would incur the
      same penalty in terms of the gap as the removal of false negatives ($\omega=2$), and require $\eps > 4\delta$
      for BMoM to operate.
    }\label{fig:ifpp-ub-lb}
\end{figure}

Both DFE-C and BMoM require space to reduce the probability of mistakes caused by statistical noise
($\gamma_\good$, $\gamma_\bad$), but only BMoM requires space ($\alpha$ and~$\omega$) to reduce the probability of
``intrinsic'' mistakes caused by an unlucky choice of the basis of the stabilizer group.  Where DFE-C partitions
the interval $\ltoI\delta,\eps\rtoI$ as
\begin{equation*}
  \delta < \delta\cdot(1+\gamma_\good) = \beta = (1-\gamma_\bad)\cdot\eps < \eps
\end{equation*}
for BMoM the partition is
\begin{equation}\label{eq:threshold-equation}
  \delta < \omega\delta < \omega\delta\cdot(1+\gamma_\good)
  =
  \beta
  = (1-\gamma_\bad)\cdot\alpha\eps
  < \alpha\eps
  < \eps.
\end{equation}
While DFE-C requires $\delta<\eps$ in order to be able to certify, BMoM --- with the analysis presented in this
paper --- works with $\omega = 2$ and, for example, $\alpha = 1/4$ if $n\ge 47$ (Fig~\ref{fig:ifpp-ub-lb} or cf.\
\S\ref{FULL::sssec:techout:falsepos:mntcarlo} of the full paper).

\paragraph{The correctness of BMoM,}\label{par:BMoM:bmom:param:correct}%
i.e., low probability of making mistakes, rests on the following two facts the details of which will be expanded in
the two Sections \ref{FULL::sec:techout} and~\ref{FULL::sec:proofs} of the full paper:
\begin{enumerate}[label=(\alph*)]
\item\label{enum:BMoM:bmom:param:correct:fneg}%
  No false negatives: For every good state~$\varrho$, with probability~$1$ over the choice of basis\footnote{The
    reference to the basis here is redundant: Every element of every basis $=$ every non-identity stabilizer.  The
    formulation was chosen to have symmetric statements for false negatives/positives.} of the stabilizer group,
  every element~$x$ of the basis satisfies $\trip{x}{\varrho} \ge 1-\omega\delta$ with $\omega := 2$.
\item\label{enum:BMoM:bmom:param:correct:fpos}%
  Few false positives: There is a function $\eta_n(\alpha)$, such that for every bad state~$\varrho$, with
  probability $1-\eta_n(\alpha)$ over the choice of the basis of the stabilizer group, at least one of the bases
  elements, $x$, satisfies $\trip{x}{\varrho} \le 1-\alpha\eps$.  We call $\eta_*(\centerdot)$ the worst-case
  intrinsic false-positive probability, and prove upper and lower bounds (cf.\
  \S\ref{FULL::sssec:techout:falsepos:mntcarlo} of full paper).
\end{enumerate}

To derive~$m$, one must first decide on an~$\alpha$, and then proceed in the same manner as for DFE-C, except that
excluding false negatives requires a union bound over the elements of the basis: As the minimum is taken, the
estimates of the expectation values of the basis elements must \emph{all} be good enough.  This incurs a
multiplicative factor of~$n$ on the probability accounted for statistical noise towards false negatives, and hence
an additive penalty of $\ln n$ on the required number of inner loop iterations (in each of the~$n$ outer loop
iterations), independently of~$\gamma_\good$.

\paragraph{About the constant~2.}\label{par:BMoM:bmom:param:2}  %
As just said, the factor $\omega=2$ rules out intrinsic false-negatives deterministically.

% For intrinsic false positives, we first find a value for~$\alpha$ that leads to deterministic absence of false
% positives (see \S\ref{sssec:techout:falsepos:determini}) but consider that factor to be too bad, and proceed to
% search for~$\alpha$ that result in absence of false positives only with high probability over the choice of the
% basis.

It seems plausible that the factor~$2$ % that excludes intrinsic false negative deterministically
can % also
be replaced by a factor $\omega < 2$ to exclude false negatives only with high probability over the choice of the
basis.  Indeed, the attentive reader of the full paper will find, in Section~\ref{FULL::sec:proofs} there, that the
two intrinsic probabilities both arise from a random variable with expectation $1/2+o(1)$, and false positives
require bounding the lower tail probability while false negatives would require to bound the upper tail
probability.

Improving the factor~$2$ might lead to an improvement in the BMoM method, as, e.g., the saved space in the interval
$\ltoI\delta,\eps\rtoI$ could be accounted towards the $\gamma_*$'s, and so reduce the number of shots.  In this
paper, we're content with $2 = O(1)$.

\subsubsection{Numbers of shots/iterations in DFE-C vs BMoM}\label{sssec:BMoM:bmom:iters}
In terms of the \emph{total} number of shots for BMoM vs DFE-C, as mentioned in Tab.~\ref{tab:dfe-M--vs--bmom-m}
above, already the number~$m$ of inner-loop repetitions of BMoM will be higher than the number~$M$ of repetitions
with DFE-C, as (a) the available gap is smaller --- $\alpha\eps - 2\delta$ in BMoM vs $\eps-\delta$ in DFE-C ---
and (b) we require a union-bound as explained above.

In short, unless either shots are \emph{considerably} less expensive than loops that involve modifying the
observable, or the distinction of the two loops gives additional information, BMoM sucks.

% ChatGPT: Attached the introduction section of a draft manuscript that I plan to submit to the journal "Quantum
% Information & Computation" (Sciendo).  Please proof-read it.  I'm *not* interested in small stylistic
% improvements, only downright mistakes.

No clear instances of outright grammatical errors or significant misuses of language appear in this text.  It is
well-structured and uses consistent, correct grammar.  Aside from potential minor stylistic changes (which you
indicated you do not want), there are no noticeable mistakes that need correcting.

%%%%%%%%%%%%%%%%%%%%%%%%%%%%%%%%%%%%%%%%%%%%%%%%%%%%%%%%%%%%%%%%%%%%%%%%%%%%%%%%%%%%%%%%%%%%%%%%%%%%

\section*{Acknowledgments}
This research was funded by Horizon Europe through the Quantum Flagship project ``OpenSuperQPlus100'',
ID~101113946.  The author also acknowledges support from the Estonian Ministry of Education and Research through
the Center of Excellence ``Foundations of the Universe'', TK202U7.

%%%%%%%%%%%%%%%%%%%%%%%%%%%%%%%%%%%%%%%%%%%%%%%%%%%%%%%%%%%%%%%%%%%%%%%%%%%%%%%%%%%%%%%%%%%%%%%%%%%%%%%%%%%%%%%%
\bibliographystyle{plain}

%%%%%%%%%%%%%%%%%%%%%%%%%%%%%%%%%%%%%%%%%%%%%%%%%%%%%%%%%%%%%%%%%%%%%%%%%%%%%%%%%%%%%%%%%%%%%%%%%%%%%%%%%%%%%%%%
\end{document}

%% file: defsN.amsart.tex
\theoremstyle{plain}

\newtheorem*{theorem*}{Theorem}

\newtheorem*{proposition*}{Proposition}

\newtheorem*{corollary*}{Corollary}

\newtheorem*{lemma*}{Lemma}

\newtheorem*{observation*}{Observation}

\theoremstyle{definition}

\newtheorem*{definition*}{Definition}

\newtheorem*{remark*}{Remark}

\newtheorem*{remarks*}{Remarks}

\newtheorem*{notation*}{Notation}

\newtheorem*{example*}{Example}

\newtheorem*{exercise*}{Exercise}

\theoremstyle{remark}

\newtheorem*{claim*}{Claim}

\newtheorem*{conjecture*}{Conjecture}

\newtheorem*{question*}{Question}

\newtheorem*{questions*}{Questions}

\newtheorem*{problem*}{Problem}

\newtheorem*{problems*}{Problems}

\newtheorem*{openproblem*}{Open Problem}

\newcommand{\subclass}[1]{}

\newcommand{\enumTi}[1]{\renewcommand{\theenumi}{#1}}

\newcommand{\alphenumi}{\enumTi{\alph{enumi}}}

\newcommand{\romenumi}{\enumTi{\roman{enumi}}}

%% file: defs.common.tex
\newlength{\hspaceforlengthglumpf}

\renewcommand{\em}{\sl}

\DeclareMathOperator{\tr}{tr}

\newcommand{\lt}{\left} 
\newcommand{\rt}{\right}

   \newcommand{\ltoI}{\left]}
\newcommand{\rtI}{\right]}  \newcommand{\rtoI}{\right[}

\newcommand{\nfrac}[2]{{\nicefrac{#1}{#2}}}

\newcommand{\ZZ}{\mathbb{Z}}

\newcommand{\eps}{\varepsilon}

\newcommand{\bra}[1]{{\langle #1 \rvert}}
\newcommand{\ket}[1]{{\lvert  #1 \rangle}}

\newlength{\algotabbingwidth}